\documentclass[twocolumn]{aastex631}

\newcommand{\hi}{\mbox{H\,{\sc i}}}
\newcommand{\hii}{\mbox{H\,{\sc ii}}}

\graphicspath{{./}{figures/}}

\accepted{June 7, 2022}

\shorttitle{$\Omega_{{\rm bar}}$ in NGC 5597}
\shortauthors{Garcia-Barreto \& Momjian (2022)}

\begin{document}
\title{$\Omega_{{\rm bar}}$ in NGC 5597 from VLA \hi\ 21 cm Observations }

\correspondingauthor{J. Antonio Garcia-Barreto}
\email{j.antonio.garcia.barreto@astro.unam.mx}

\author[0000-0002-3773-9613]{J. Antonio Garcia-Barreto}
\affiliation{Instituto de Astronomia,
Universidad Nacional Aut\'onoma de M\'exico, \\
Apartado Postal 70-264, Ciudad de M\'exico 04150, M\'exico}

\author[0000-0003-3168-5922]{Emmanuel Momjian}
\affiliation{National Radio Astronomy Observatory,
P. V. D. Science Operations Center, \\
P.O. Box O, 1003 Lopezville Road, Socorro, 87801, New Mexico, U.S.A.}

\begin{abstract}
We report Very Large Array B-configuration observations of the atomic hydrogen 21 cm line emission from the barred disk galaxy NGC 5597 at an angular resolution of $\sim 7\farcs1 \times 4\farcs2$. Using the resonance method, and assuming the ratio of the corotation radius to the semi-major axis of the stellar bar is unity ($\mathcal{R} \equiv R_{\rm CR}/a_{\rm bar} = 1$), we estimate the angular pattern speed of the stellar bar to be, $\Omega_{\rm bar} \sim 15.3$\,km\,s$^{-1}$\,kpc$^{-1}$. This constant value for $\Omega_{{\rm bar}}$ crosses $\Omega_{{\rm gas}} + \kappa(R)/4$ at a distance $\sim 6.73$ kpc which would correspond to the spatial location of the north spiral structure near an outer m=4 resonance. This value of $\Omega_{\rm bar}$ is similar to the values estimated for other bright nearby barred galaxies that exhibit circumnuclear rings (near ILR) or outer rings (near OLR).
\end{abstract}

\keywords{galaxies: active: individual (NGC 5597) ---galaxies: kinematics and dynamics --- galaxies: ISM --- galaxies: spiral}

\section{Introduction} \label{sec:intro}

    There is growing evidence for the existence of weak nuclear activity in normal
disk galaxies, with and without a prominent stellar bar, supported by observational detections of nuclear H$\alpha$ low velocity bipolar outflows, e.g., M81 Sb(r)I-II \citep{goa76}, NGC 1068 Sb(rs)II \citep{ulv87} which was more recently reported to host a stellar bar \citep{sco88b}, M51 Sbc(s)I-II \citep{for85,cec88,cra92,sco98}, NGC 3079 Sc pec: \citep{vei94}, M101 Sc(s)I \citep{moo95}, NGC 3367 SBc(s)II \citep{gar98,gar02}, and NGC 1415 SBa late \citep{gar19}. This suggests that such galaxies represent a low end of scale for nuclear activity after quasars, BL Lac objects, radio galaxies, and Seyfert galaxies. Such weak nuclear activity in normal disk galaxies originates around central massive black holes ($M_{\rm BH} \sim$ a few $10^6 M_{\odot}$), with the level of the activity being governed by the gas supply to fuel these central engines \citep{nor83}. An open question in the case of barred galaxies is whether there is a correlation among the onset of nuclear activity, their central gas, and the angular velocity pattern of their stellar bar.

    Stellar bars\footnote{Sustained by stars in $x_1$ orbits
that lie along the major axis of the bar \citep{con80a,con80b}} represent a small non-axisymmetrical gravitational potential {\bf $\Phi_{\rm {bar}}$} compared to the axisymmetric gravitational potential of a disk galaxy {\bf $\Phi_{\rm^{disk}}$}. 

    NGC 5597, located in the Southern Virgo - Libra Cloud of Galaxies, is one such
nearby galaxy with a prominent stellar bar. It is a bright (m$_B \sim 12.57$) disk galaxy classified as SBc(s) II \citep{san87} and as SAB(s)cd in NED\footnote{The NASA/IPAC Extragalactic Database (NED) is operated by the Jet Propulsion Laboratory, California Institute of Technology, under contract with the National Aeronautics and Space Administration.}. It forms a gravitationally bound isolated pair with the disk galaxy NGC\,5595~(Sc). 

    NGC 5597 is one of our original set of 56 bright and nearby Shapley-Ames barred
galaxies with IRAS colors that suggest central massive star formation. Their 2D optical images in the red continuum filter I (8040 \AA) and in the H$\alpha$+N II narrow line filter have been published by \citet{gar96}. The IRAS 60$\mu$m and 100$\mu$m flux densities of NGC 5597 are 8.7\,Jy and 15.32\,Jy, respectively, with a dust temperature of $T_{\rm dust} \sim 36$\,K. The (IRAS) far-IR luminosity is $L_{\rm FIR} \sim 2.2 \times 10^{10} L_{\odot}$ \citep{gar96}.

    In this paper, we present Karl G. Jansky Very Large Array (VLA) observations of
the atomic hydrogen, \hi\ 21 cm, gas emission from the barred disk galaxy NGC 5597 to determine its two dimensional gas velocity field and estimate its bar's angular pattern speed, $\Omega_{\rm bar}$. For this measurement, we utilize the resonance method that has been proven to be a valid technique to apply on nearby bright barred galaxies with rings near the Inner Lindblad Resonance (ILR) or Outer Lindblad Resonance (OLR). The outline of this paper is as follows: Section 2 presents the observations, data reduction, and analysis, Section 3 reports the results and discussion, and Section 4 presents the summary and conclusions.

\section{Observations and Data Reduction} \label{sec:observations}

    The observations to image the \hi\ 21 cm emission from the barred disk galaxy NGC
5597, and map its two dimensional velocity field, were carried out with the VLA of the NRAO\footnote{The National Radio Astronomy Observatory is a facility of the National Science Foundation operated under cooperative agreement by Associated Universities, Inc.} in its B-configuration (maximum baseline $b_{max}=11.1$ km) on 2019 June 6, 7, 11, 16, 18, \& 22.

    Each of the six observing sessions had a duration of 1.67\,hr (100\,min on-source
time). The L-band receiver of the VLA was tuned to the redshifted frequency of the \hi\ 21 cm line ($\nu_{\rm rest} = 1420.405752$ MHz) at a heliocentric velocity\footnote{NGC 5597 was observed simultaneously with the nearby galaxy NGC 5595 using a single pointing midway between the two sources. They have very comparable recession velocities of $\sim 2700$ km s$^{-1}$} corresponding to $V = 2700$ km s$^{-1}$.  (All velocities in this paper are heliocentric, using the optical convention for redshift.) The Wideband Interferometric Digital ARchitecture (WIDAR) correlator of the VLA was set to deliver a single 4 MHz subband with 256 spectral channels ($\Delta\nu_{\rm channel} = 15.625$ kHz), covering a velocity range of $\Delta V \sim 843$ km s$^{-1}$. This velocity span is sufficient to cover the full width at zero intensity of the \hi\ 21 cm emission line from NGC 5597, and provide line-free channels on both sides of the \hi\ emission for continuum subtraction. In addition to the target field, we observed the source 3C\,286 as the flux density scale and bandpass calibrator, and J1448-1620 as the complex gain calibrator. A more detailed description of the system setup will be provided in a forthcoming paper on the \hi\ 21 cm kinematics and dynamics of both NGC\,5595 and NGC 5597 (Garcia-Barreto \& Momjian 2022, in preparation).

    We utilized both the Common Astronomy Software Applications (CASA) and the
Astronomical Image Processing System (AIPS) packages of the NRAO for data reduction and analysis. The initial calibration steps, e.g., bandpass, flux density scale, complex gain calibration, continuum subtraction, and imaging, were performed in CASA. All the spectral and kinematics analysis were carried out in AIPS.

\begin{deluxetable*}{lllcllcccc}[t]
\tablecaption{NGC 5597: Coordinates, Hubble type, Distance, Systemic \hi\ Velocity}  \label{tab:table} 
\tablehead{
\colhead{Galaxy} & \colhead{$\alpha$(J2000)} & \colhead{$\delta$(J2000)} & \colhead{Ref.} & \colhead{RSA type} & \colhead{NED type} & \colhead{Distance} & \colhead{Ref.} & \colhead{$V({\hi})_{\rm sys}$} & \colhead{Ref.} \\
\colhead{Name}  & \colhead{$hh~mm~ss$}    & \colhead{$\degr~~\prime ~~\farcs$} & \colhead{} & \colhead{} & \colhead{} & \colhead{Mpc}  & \colhead{} & \colhead {km s$^{-1}$} & \colhead{} }
\colnumbers
\startdata
NGC 5595 &  $14~24~13.3$  & $-16~43~21.6$ & 1 & Sc(s) II  &
SAB(rs)c & 38.6 & 2 & 2697 & 3 \\
NGC 5597 &  $14~24~27.49$ & $-16~45~45.9$ & 1 & SBc(s) II & SAB(rs)b & 38.6 & 2 & 2698 & 3\\
\hline 
\enddata
\tablecomments{1) \citep{dia09}, 2) \citep{tul88}, 3) \citep{pat03} }
\end{deluxetable*}

\begin{figure*}[bht]
\includegraphics[width=15cm,height=14cm]{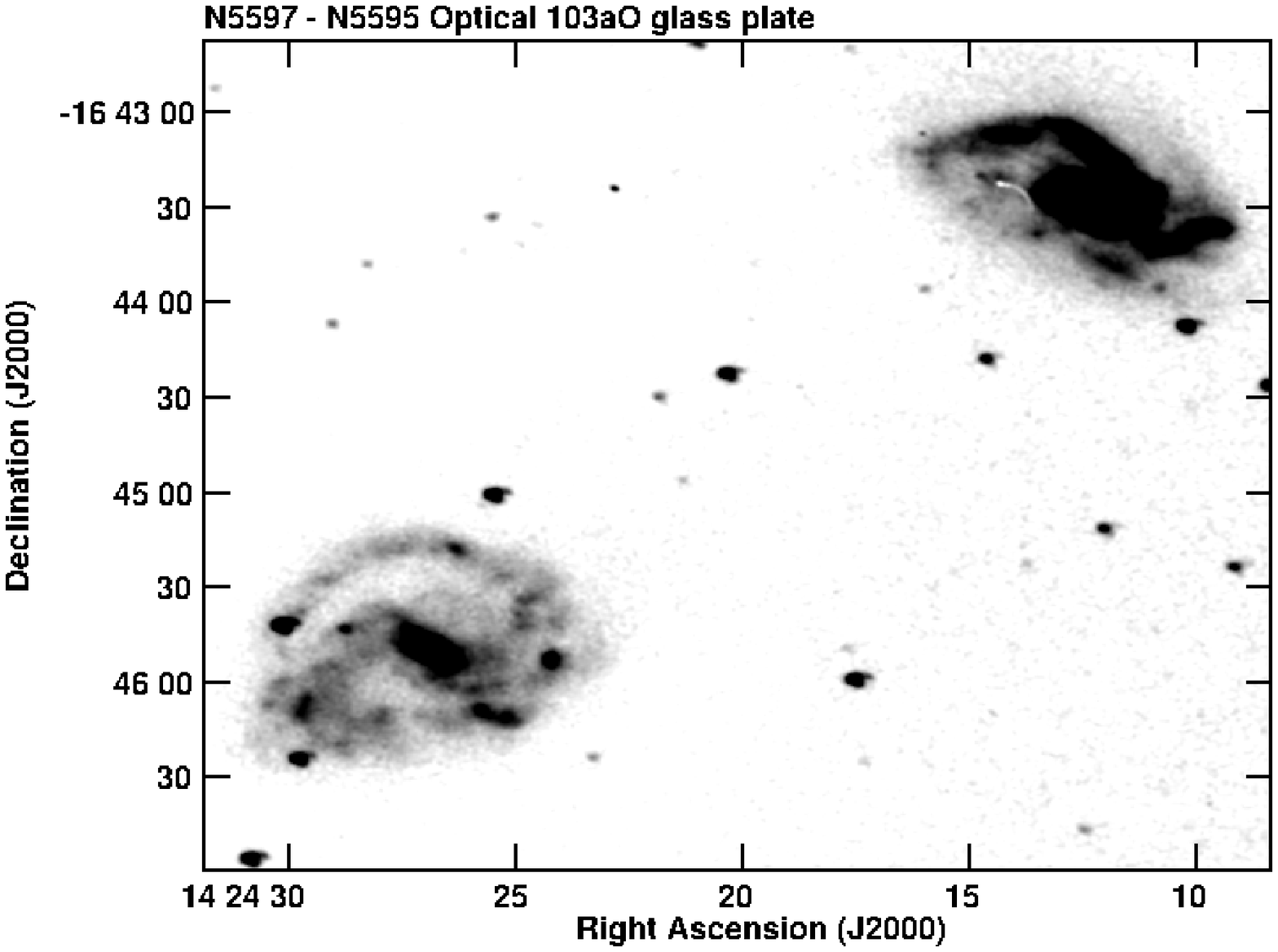}
\figcaption{Reproduction of the optical blue continuum image (glass plate 103aO) of the close-pair disk galaxies NGC 5597 at the south east and NGC 5595 at the north west, obtained with the OAN-SPM 2.1\,m optical telescope in Mexico \citep{dia09}. The flux density scale is not calibrated. Grey scale is from $2 \rightarrow 27.8 \sigma$, where $\sigma$ is the noise in arbitrary units.  \label{fig. 1}}
\end{figure*}

    Image cubes from each of the six observing sessions were made and examined to
determine the \hi\ line and the line-free channels. Then the continuum emission was subtracted in the $uv$ plane. The final image cube using the combined $uv$ data of all the sessions was produced with a cell size of 1$\farcs0$, and Briggs weighing with robust parameter value of 0.8 in CASA, resulting in a synthesized beam size, at full width half maximum (FWHM), of $\sim 7\farcs1 \times 4\farcs2$ (P.A.$\sim -10\degr$). This angular resolution corresponds to a linear resolution of $\sim 1.33 \times 0.78$ kpc$^2$. 
    The final \hi\ image cube was imported to AIPS for further analysis. In there
Moments 0, 1, and 2 were made using a flux density cutoff of 2.5$\sigma$, with 1$\sigma \sim 450\,\mu$Jy\,beam$^{-1}$\,channel$^{-1}$.

\normalsize
\section{Results and Discussion}
\subsection{NGC 5597: an SBc disk galaxy}

    The disk galaxies NGC 5595 and NGC 5597 are in the Southern Virgo - Libra Cloud of
galaxies, Tully's group 41-14(+14), with a very low galaxy volume density of only 0.16 galaxies Mpc$^{-3}$; see their spatial location at the galactic coordinates $l^{II} \sim 332\rlap{.}{\degr}8$, $b^{II} \sim +40\rlap{.}{\degr}7$ in  Plates 1 and 5 in \citet{tul87}, and in $\alpha$ and $\delta$ in Fig.\,2 of \citet{tam85}. Since these two galaxies are separated by $\Delta\alpha \sim 13\rlap{.}{^s}54$, $\Delta\delta \sim 138\arcsec$, or about $3\rlap{.}{'}97$ on the plane of the sky while having similar systemic velocities (see Table 1), we may confidently say that they form a physically isolated and gravitationally bound pair of disk galaxies. Therefore, in this paper, we adopt a Hubble spectroscopic distance of $D_{{\rm pair}} = 38.6$ Mpc \citep{tul88} giving an angular-to-linear scale conversion of $1\arcsec = 187.14$ pc.

    Figure 1 shows a reproduction of the optical blue continuum 103aO emission from
the pair of disk galaxies NGC 5597 (south-east) and NGC 5595 (north-west) in grey scale (relative units) taken with the OAN-SPM 2.1\,m optical telescope in Mexico \citep{dia09}. In this blue optical continuum image, there are no bridges or structures connecting the two galaxies. Figure 2 shows a reproduction of the optical red, filter I 8040 \AA, continuum image of NGC 5597 convolved with a Gaussian beam of $\sim 1\farcs5 \times 1\farcs5$ at FWHM \citep{gar96}. The grey scale and the contours in this figure are also in relative units. Both the blue and red optical continuum images of NGC 5597 show a central elongated structure extending about $2a \sim 28\farcs3$ by $2b \sim 14''$ at a P.A. $\sim 52\degr$. We will refer to this structure as the stellar boxy bar of NGC 5597 (see the south-east galaxy, NGC 5597, in Figure 1, and its more detailed view in Figure 2).

    Additionally, in Figure 1, and more specifically in Figure 2, there are at least
four narrow and curved structures as spiral arms outside the central region in NGC 5597 (labeled 1 to 4 in Figure 2). From the optical red continuum image, the first structure (1) starts from the NE half of the boxy stellar bar and extends to the SE, the second structure (2) starts from the SW of the southern half of the boxy bar and extends further to the SW, the third structure (3) starts from the NW of the southern half of the boxy bar and extends further NW joining the outer fourth structure. The fourth structure (4) seems to start from the SW of the disk and continues counter clockwise to N, NE, and SE just parallel on the outside of the first structure.\\

\begin{figure*}[bht]
\includegraphics[width=15cm,height=14cm]{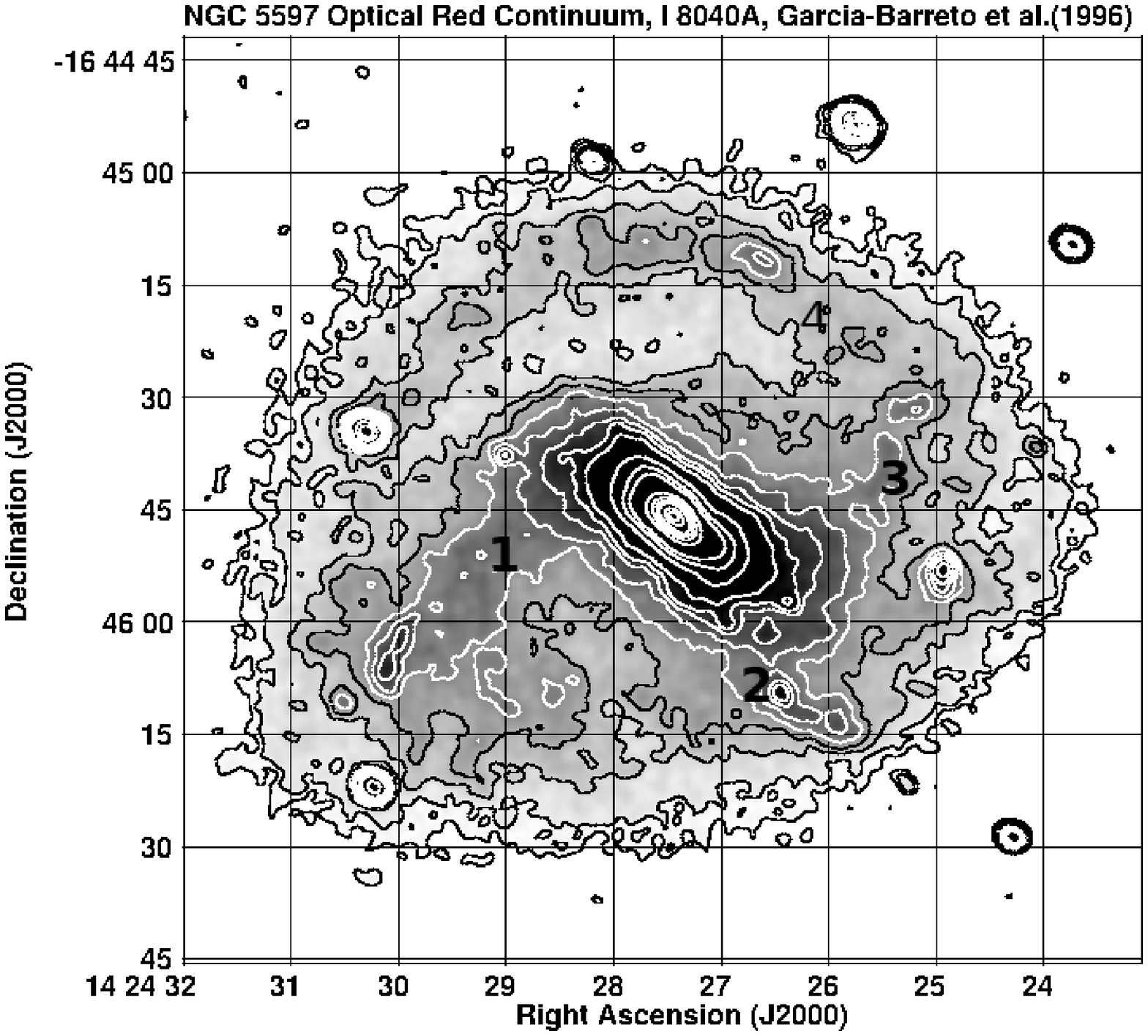}
\figcaption{Reproduction of the optical red, filter I 8040 \AA, continuum image that has been convolved with a Gaussian beam FWHM $\sim 1\farcs5 \times 1\farcs5$ in grey scale and in contours \citep{gar96}. The image had not been calibrated in intensity scale, thus the grey-scale stretch is in relative units proportional to noise from 5$\sigma$ up to 37.7$\sigma$, where 1$\sigma = 45$ in arbitrary units. Contours are 5, 7, 11, 15, 20, 25, 30, 34, 40, 60, 80, 100, 200, 300, 400, 500, 750, 900 $\times 1\sigma$. Note the boxy isophotes at radii larger than $5\farcs0$ that delineate a structure straddling the nucleus elongated in P.A.$\sim 52\degr$ EofN.
\label{fig. 2}}
\end{figure*}

\subsubsection{\bf Estimation of Stellar Bar in NGC 5597}

     In general, the estimation or the determination of the size of a stellar bar, even for
nearby bright disk galaxies, has been a challenging topic for not only observational studies, but also for image analysis as well as computer simulations \citep{erw05}. From the observational perspective, some of the reasons are: the wavelength of the continuum images -- optical blue, red, IR, MIR --, inclusion of the bulge, subtraction of intensity from the bulge, and the no-agreed-upon method --the projected semimajor axis of the stellar bar on the plane of the sky, the projected semimajor axis of the stellar bar on the kinematical semimajor axis of the disk galaxy, and the radius where the intensity along the stellar bar has fallen to half of its peak value.
    From the computer image analysis perspective, some of the reasons are: the different and
elegant mathematical algorithms -- ellipse fitting, fourier transform m=2 amplitudes, modified Ferrer profiles, and Lagrangian radius, and from the computer N-body simulations perspective: the angular momentum transfer, gas viscosity, galaxy harassments, galaxy close-by interactions,  \citep{kor79,ken90,con80a,con80b,qui94,mar95,deb00,ath02,ath03,erw05,sal15,dia16}.

    Kormendy (1977, 1979, 1983) did extensive work in modeling different
independent stellar components in galaxy brightness profiles. For instance, he identified the most fundamental components to be: spherical central bulge, exponential axisymmetric disk, and bar, and as secondary components: lens, ring, outer ring, and outer lens. He also concluded that the total mass of a disk galaxy does not determine whether the galaxy makes a bar, but if it does make one, the total mass uniquely determines the size of the bar \citep{kor79}. \citet{ken89} and \citet{ken90} 
also made detailed photometry of two barred galaxies, NGC 936 and NGC 4596, with analytic mathematical expressions to the luminosity profiles of each component (bulge, bar, and disk) to construct bar models. They did give two values for $a_{\rm bar}$ for each galaxy: i) on the plane of the sky and ii) projected on the major axis of the disk. In particular,  $a_{\rm bar}$ in NGC 4596 was reported to be the radius where the intensity along the major
axis of the bar has fallen to half of its peak value \citep{ken90}.

    For NGC 5597, previous estimation of its stellar bar's semi-major axis
by the same group of astronomers has been published with surprisingly two different values, namely, $a_{\rm bar} \sim 3\farcs79$ \citep{dia16}, and $a_{\rm bar} \sim 6\farcs51$ \citep{sal15}\footnote{Salo et al. 2015 from 3.6 $\mu$m S$^4$G imaging, used a modified Ferrers profile to characterize bars, while Diaz-Garcia et al. 2016 from the same 3.6 $\mu$m S$^4$G imaging used i) maximum of the tangential to radial force, ii) m=2 Fourier density amplitude and iii) isophotal ellipticity, to characterize bars.}. 

    Figures 3 and 4 (left panels) show the innermost regions of the blue (103aO)
and red (8040 \AA ) images, extracted from Figures 1 and 2, to provide a more detailed view of the elongated stellar structure in NGC 5597, notice the lack of dust lanes in both images. The right panels in Figures 3 and 4 show the 1-D slice with intensity versus angular distance along P.A.$\sim 52\degr$ EofN. Notice the exponential profile in the slice of the red image (Figure 4). 
    The full width at half maximum intensity of the blue and red image profiles are FWHM$_{\rm
blue} \sim 10\farcs3$ and FWHM$_{\rm red} \sim 5\farcs1$, respectively.

\begin{figure*}[t!]
\includegraphics[width=8cm,height=8cm]{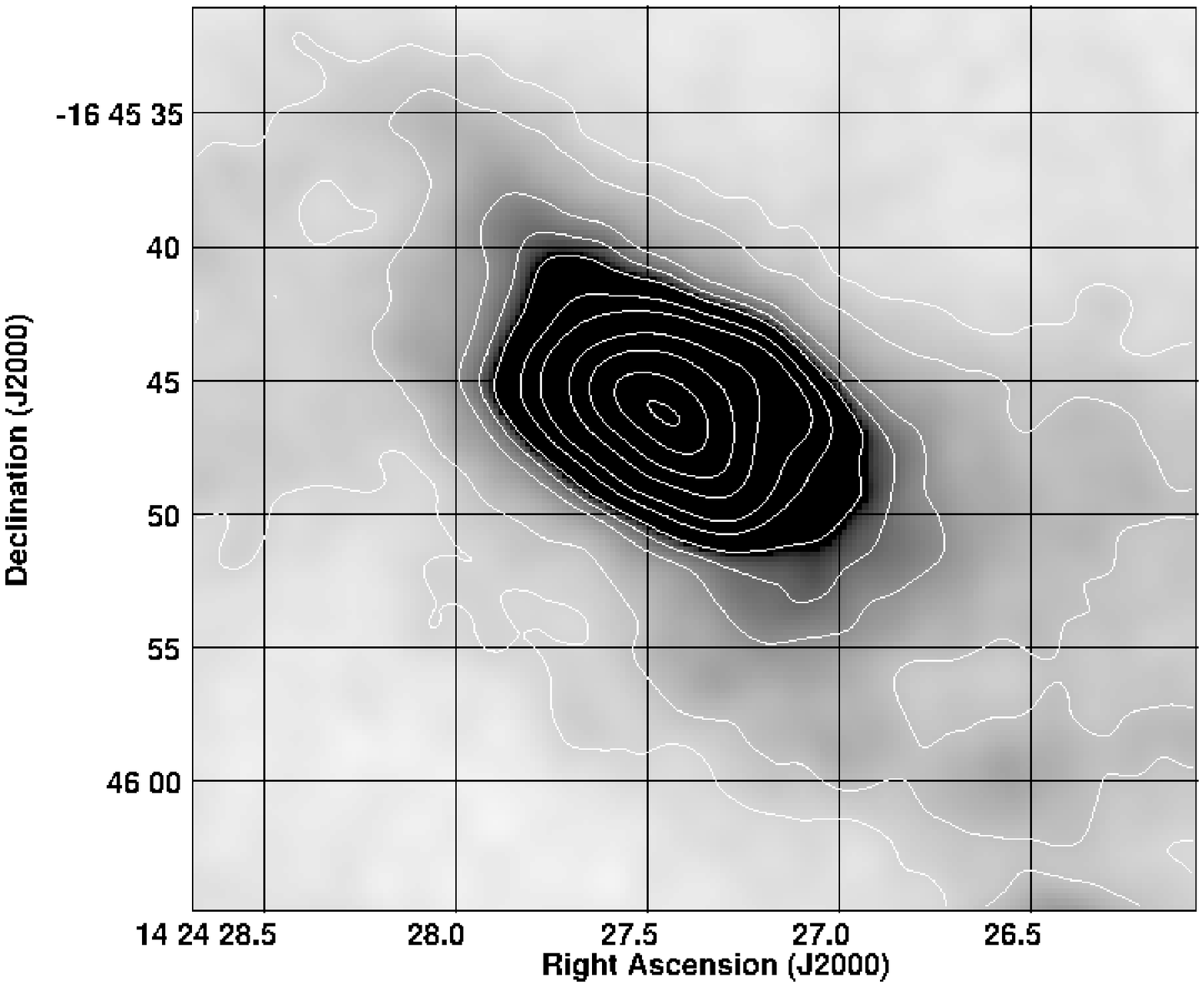}
\includegraphics[width=8cm,height=8cm]{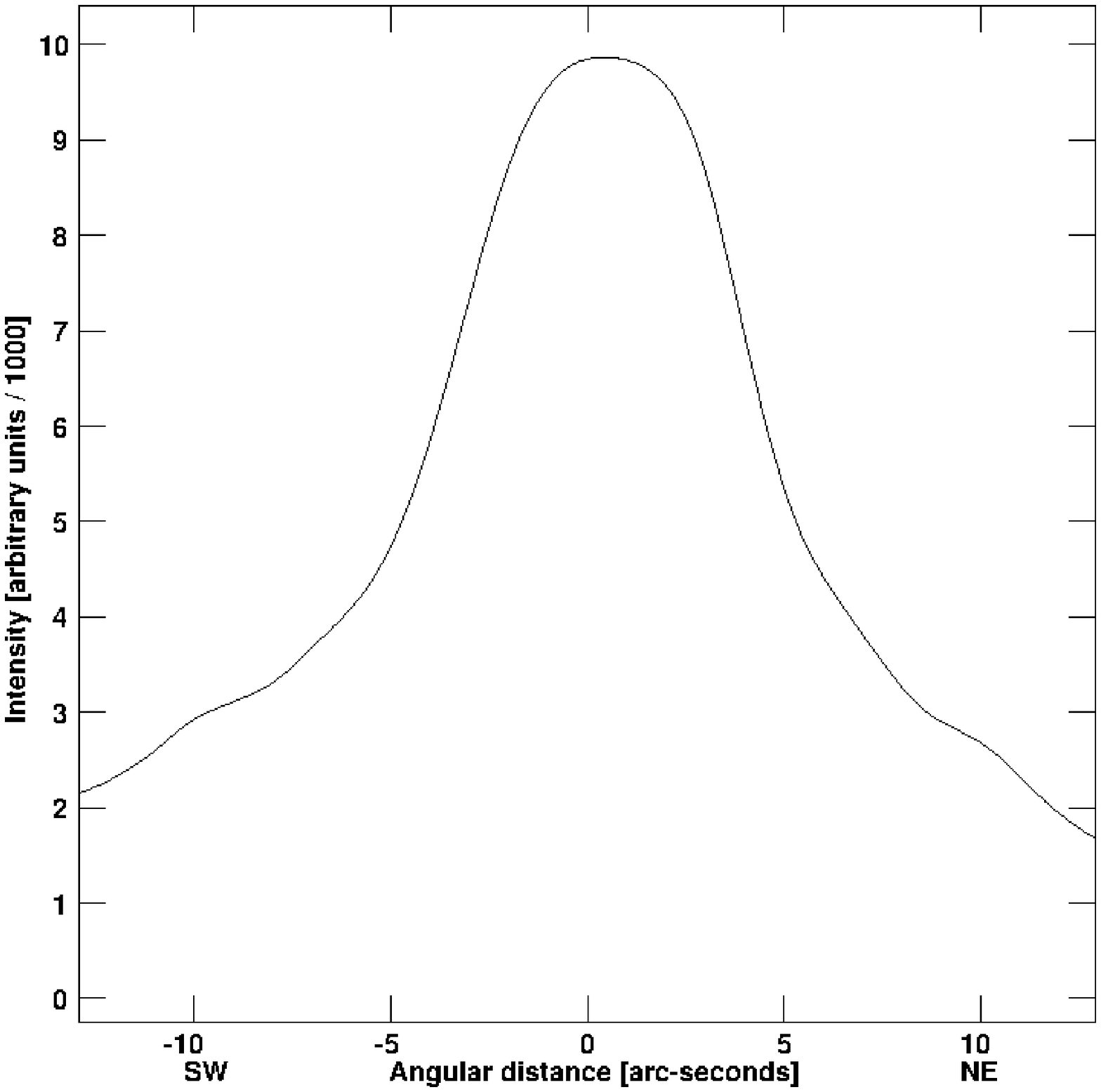}
\figcaption{Left panel: innermost optical blue, 103aO, continuum image of the stellar bar in NGC 5597 in grey scale and in contours, extracted from Figure 1 \citep{dia09}. The image had not been calibrated in intensity scale, thus the grey-scale stretch is in relative units proportional to noise from 1.5$\sigma$ to 39$\sigma$, where 1$\sigma = 90$ in arbitrary units, and the contours are at 15, 20, 30, 35, 40, 50, 60, 80, 95, 107, 111.5 $\times 1\sigma$, showing the extent of the boxy isophotal contours before a contour with azimuthal extensions. Boxy isophotes delineate a structure straddling the nucleus elongated in P.A.$\sim 52\degr$ EofN. Right panel: one dimensional slice with intensity versus angular distance at P.A.$\sim 52\degr$ EofN. Notice the lack of dust lanes.
\label{fig. 3}}
\end{figure*}

\begin{figure*}
\includegraphics[width=8cm,height=8cm]{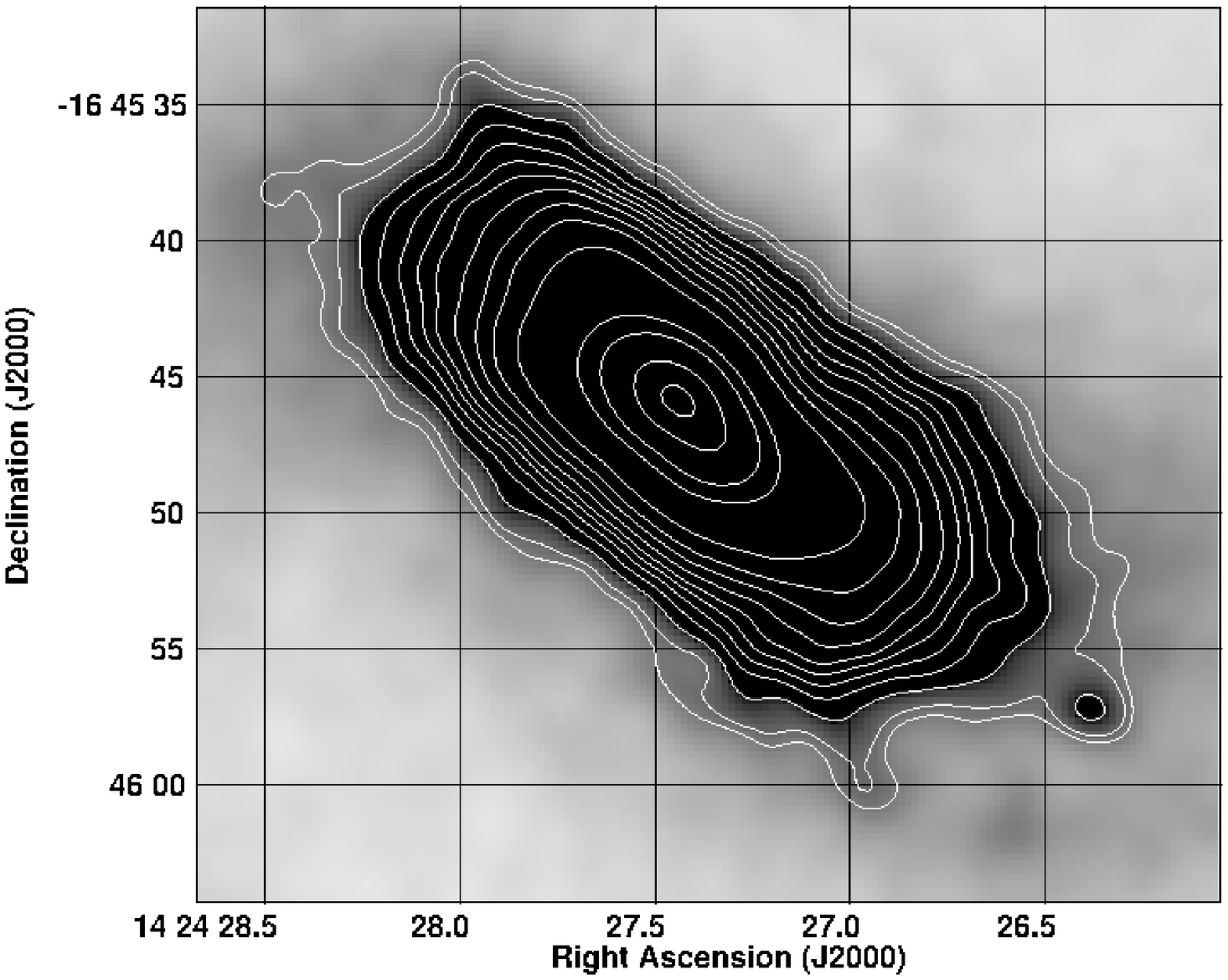}
\includegraphics[width=8cm,height=8cm]{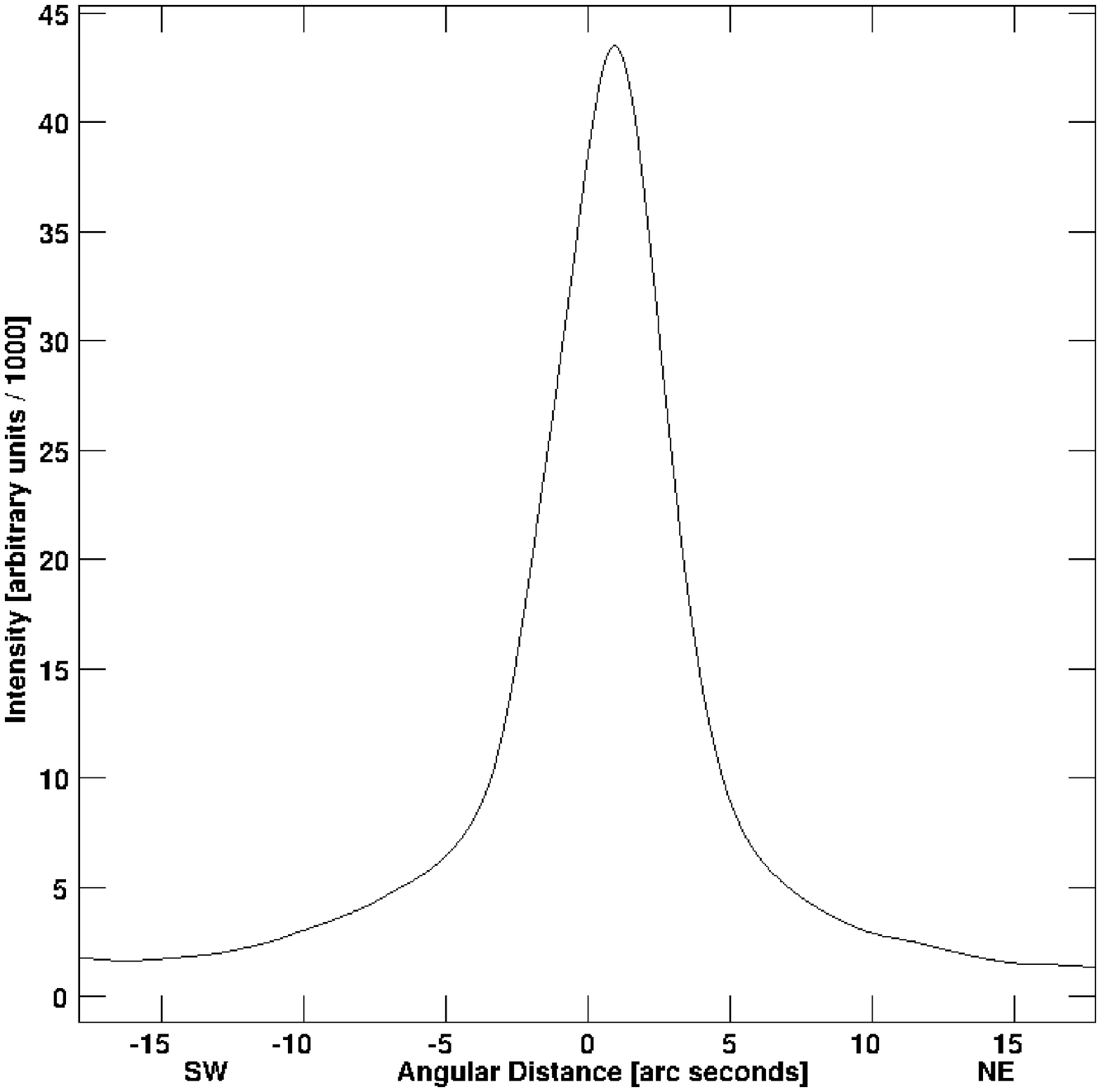}
\figcaption{Left panel: innermost optical red, filter 8040 \AA, continuum image of the stellar bar in NGC 5597 in grey scale and in contours, extracted from Figure 2 \citep{gar96}. The image had not been calibrated in intensity scale, thus the grey-scale stretch is in relative units proportional to noise from 2.8$\sigma$ to 40$\sigma$, where 1$\sigma = 45$ in arbitrary units, and the contours are at 34, 36, 40, 44, 48, 52, 56, 62, 70, 80, 100, 200, 300, 600 and 900 $\times 1\sigma$, showing the extent of the boxy isophotal contours before a contour with azimuthal extensions. Boxy isophotes delineate a structure straddling the nucleus elongated in P.A.$\sim 52\degr$ EofN. Right panel: one dimensional slice with intensity versus angular distance at P.A.$\sim 52\degr$ EofN. Notice the lack of dust lanes.
\label{fig. 4}}
\end{figure*}

     In addition to the two different values reported by \citet{sal15} and \citet{dia16}, the
left panels of Figures 3 and 4 further demonstrate the challenge of deriving a reliable value for the semimajor axis of the stellar bar in NGC 5597. Here, clearly the optical red elongated structure shown by the boxy isophotal contours results in a semimajor axis value for the bar that is larger than those previously published.

    In this paper, we used the more modest visual method analyzing the optical blue
(103aO) images shown in Figures 1 and 3 and the optical red (broad band filter I 8040\,\AA) image shown in Figures 2 and 4 to estimate the semimajor axis of the stellar bar in NGC 5597. In particular, we determined the projected length, on the plane of the sky, of the semimajor axis of the stellar bar, $a_{\rm bar}$, that is, the distance from the nucleus (or photometric center) to the isophote still showing an elliptical boxy shape, just before the next isophote that deviates into azimuthally extended structures in the blue and red images. For this, we measured the distance from the center to north-east and from the center to south-west, and took the average of the two values for both blue and red images. The estimated $a_{\rm bar}$ values differ for the blue and red images, as expected, because the associated emission trace different stars; although the OB stars are brighter, the GK stars outnumber them \citep{mih81}. The mean $a^{blue}_{\rm bar} \sim 7''$, while the mean $a^{red}_{\rm bar} \sim 14''$. Curiously $a^{blue}_{\rm bar}$ roughly agrees with Salo et al. (2015) value of $a^{salo}_{\rm bar} \sim 6\farcs51$. Furthermore, for the stellar bar in NGC 5597 (with no mass distribution model), we have estimated the radius where the intensity along its major axis falls to half of its peak value for both the blue and red images: $a^{0.5b}_{\rm bar} \sim 5\farcs1$, and $a^{0.5r}_{\rm bar} \sim 2\farcs5$.

    At this point, we have six different estimates for the semimajor axis of the
stellar bar in NGC 5597 {\it on the plane of the sky}, $a^{diaz}_{\rm bar} \sim 3\farcs79$, $a^{salo}_{\rm bar} \sim 6\farcs51$, $a^{blue}_{\rm bar} \sim 7''$, and $a^{red}_{\rm bar} \sim 14''$, $a^{0.5b}_{\rm bar} \sim 5\farcs1$, and $a^{0.5r}_{\rm bar} \sim 2\farcs5$. 

    Now, we will estimate the six different semimajor axis values of the stellar bar in 
NGC 5597 {\it projected on the kinematical major axis, hereafter ma,} from the value {\it on the plane of the sky, hereafter pls,} utilizing the mathematical expression $a^{ma}_{\rm bar} = a^{pls}_{\rm bar} \times \sqrt{cos^2(\phi) + (sin^2(\phi)/sin^2(i))}$ \citep{kor83}, where $\phi$ is the angle difference between the P.A. of the kinematical semimajor axis of the disk galaxy and the P.A. of the semimajor axis of the stellar bar, and {\it i} is the inclination of the disk of the galaxy with respect to the plane of the sky. For NGC 5597, these angles are $\phi \sim 48 \degr$ and $i \sim 36\degr$. 

    Thus, the six different values for the semimajor axis of the stellar bar in NGC
5597 {\it projected on the kinematical major axis} are:  $a^{d-ma}_{\rm bar} \sim 5\farcs42$, $a^{s-ma}_{\rm bar} \sim 9\farcs31$, $a^{blue-ma}_{\rm bar} \sim 10''$, and $a^{red-ma}_{\rm bar} \sim 20''$, $a^{0.5b-ma}_{\rm bar} \sim 7\farcs3$, and $a^{0.5r-ma}_{\rm bar} \sim 3\farcs6$. The final best estimated value will be provided in our \hi\ 21 cm kinematical analysis presented in the following subsections.


\begin{figure*}[bht]
\includegraphics[width=15cm,height=14cm]{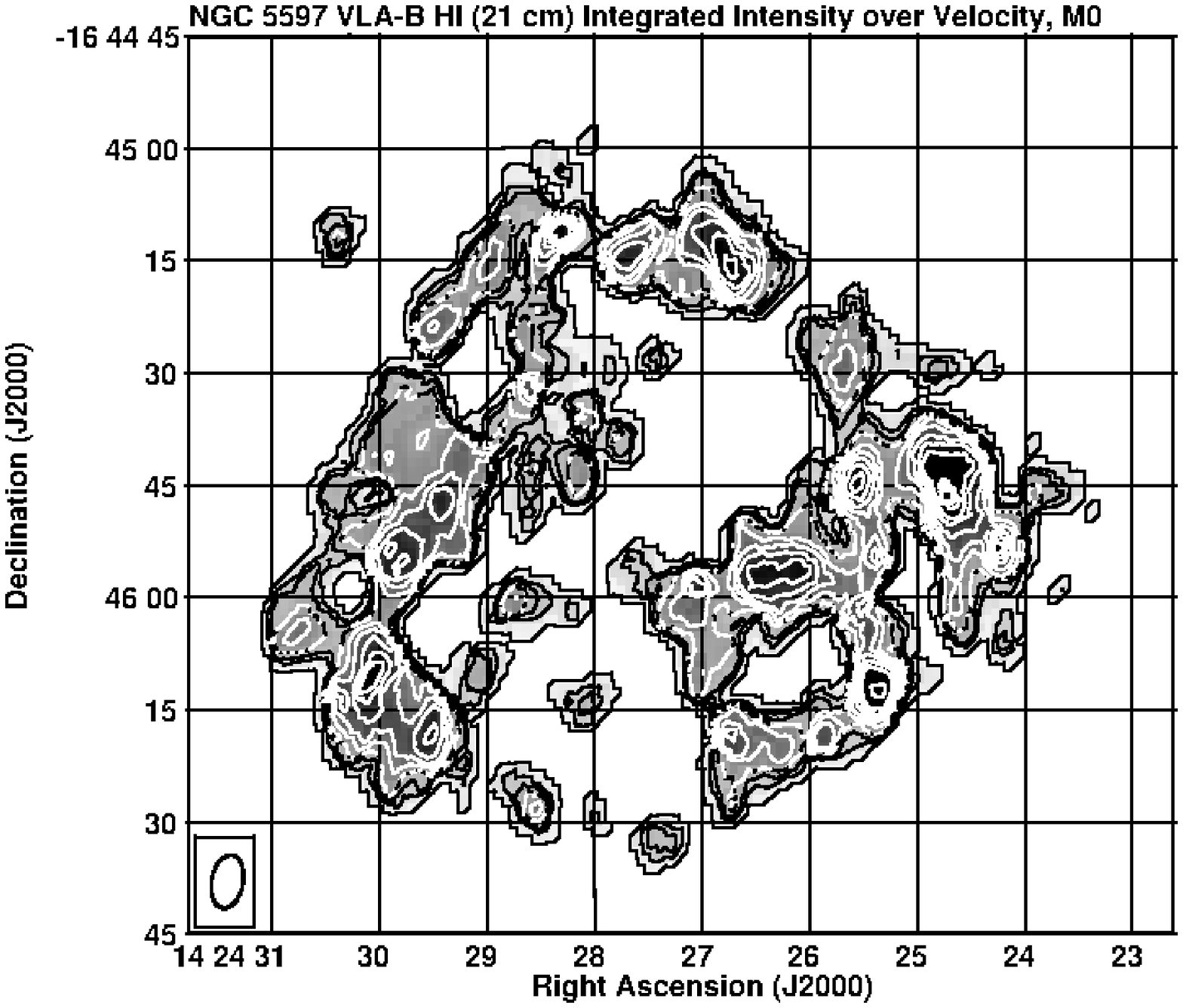}
\figcaption{VLA B-configuration \hi\ 21 cm integrated intensity over velocity, Moment 0, image of NGC 5597 in contours superposed on the same image in grey scale. Contours are 3, 35, 50, 70, 100, 125,150,175, 200, 250, 270 times $1 \sigma \sim 0.3$ mJy\,beam$^{-1}$\,km\,s$^{-1}$.
The grey scale ranges from 3$\sigma$ to 200$\sigma$. The synthesized beam ($\sim 7\farcs1 \times 4\farcs2$ at FWHM) is shown in the lower left corner.
\label{fig. 5}}
\end{figure*}

\subsubsection{VLA B-configuration \hi\ 21 cm Gas Spatial Distribution}

    The optically thin \hi\ 21 cm emission arises from cold atomic gas with an
assumed spin temperature of $T_{\rm s} \sim 100$\,K or less \citep{fie58a,fie58b}. Figure 5 shows the VLA B-configuration \hi\ 21 cm integrated intensity over velocity (Moment 0) in both contours and grey scale. At the angular resolution of these VLA observations (synthesized beam FWHM $\sim 7\farcs1 \times 4\farcs2$) all the \hi\ gas emission is within the optical disk of NGC 5597. 

        The comparison of the spatial distribution of the stars (blue and red optical
images in Fig. 1 and Fig. 2), hot gas (H$\alpha$, and cm radio continuum images, not shown here; Garcia-Barreto \& Momjian 2022, in preparation) and the cold \hi\ gas (Fig. 5) in NGC 5597 seem to indicate that the \hi\ clouds are at a slightly shorter radial distance than the few hot \hii\ regions in the northern, eastern, south-east end, and western spiral arms. In other words, new star formation is taking place in the outer edges of the spiral arms. This is the expected spatial distribution per the density wave theory of spiral arms in disk galaxies. The alternative explanation for the total cold \hi\ gas being a byproduct of photodissociation of $H_2$ regions might not seem to apply in the disk galaxy NGC 5597 \citep{all02}; there might be many photodissociation regions (PDRs) throughout the disk of NGC 5597, but they would be associated with localized star formation regions.

    In NGC 5597, there are \hii\ regions (hot gas) with no associated cold \hi\ gas. One such
example is the \hi\ hole centered at $\alpha \sim 14^h\,24^m\,26\rlap{.}^s5, \delta \sim -16\degr\,46'\,10''$, and another example is the innermost central (nucleus) region. In the innermost central region, there might exist cold dense molecular gas ($H_2$, CO) where transitions from atomic to molecular gas are favored \citep{bli06}. While there does not seem to be yet any published CO observations for NGC 5597, one may expect $M_{H_2} \sim 6.4 \times 10^9 M_{\odot}$ from the CO - FIR correlation \citep{sco88a} from its IRAS (FIR) luminosity $L_{\rm FIR} \sim 2.2 \times 10^{10} L_{\odot}$. There are few examples in the literature on other nearby bright barred galaxies with  molecular gas in their central regions, such as NGC 4314 \citep{com92}, NGC 3367 \citep{gar05,gar07}, and NGC 1068 \citep{pla91}. The innermost central region may also be the site of a nuclear bipolar outflow from a central supermassive black hole (SMBH),  similar to what has been observed in the barred galaxies NGC 1068 \citep{wil87} and NGC 3367 \citep{gar98,gar02}, or the site of a nuclear bipolar geyser as in the barred galaxy NGC 1415 \citep{gar19}. 

    As for the south-west \hi\ hole in NGC 5597 (and the H$\alpha$ image, not shown here,
Garcia-Barreto \& Momjian 2022 in preparation), it is at the approximate location of the anomalous SW spiral arm (labeled 2 in Fig. 2) with recent star formation and hot \hii\ regions. In other external disk galaxies many cold \hi\ gas holes have been detected where their large scale distribution is correlated with \hii\ regions, suggesting that the \hi\ holes are a result of intense OB star formation (i.e., M31) \citep{com02}. It might also be a site of cold dense molecular gas ($H_2$, CO), but there does not seem to be any published observations on NGC 5597 to support this.

    The NGC 5597 - NGC 5595 disk galaxy pair is in an area of the universe with very
low galaxy volume density \citep{tam85,tul87}, therefore, there is no hot intergalactic gas which could have stripped the atomic gas from outer disk radii. It seems that both NGC 5595 and NGC 5597 originated with just enough primordial atomic neutral \hi\ gas to form their disks and stars. 

    We note the narrow, curved, and long structures seen in \hi\ emission in the NE
with abrupt decrease in the emission toward the eastern side. Bright \hi\ emission arises from the inner NE spiral arm in well correspondence with the optical arm. Somewhat weaker \hi\ emission arises from the outer NE arm that seems to be at the inner edge of the optical eastern arm. The \hi\ emission from the north arm is in the inner side of the optical continuum
arm. 

         The outer northern optical spiral arm in NGC 5597 seen in Figures 1 and 2
(labeled 4 in Fig. 2), which indicates the spatial location of stars, is located at an approximate angular distance of $37'' \pm 5''$ from the nucleus (corresponding to a radial distance {\it on the plane of the sky} of about $6.9 \pm 0.94$ kpc). It is a well ordered, narrow, and curved structure seen from the north-west going counterclockwise to the north, north-east and east. The distance of the mean peak of its \hi\ emission is about $30\farcs8 \pm 5\farcs8$ or at about $5.8 \pm 1$ kpc, and the far outer edge of the atomic gas emission ends abruptly at about 7.4 kpc. This spiral arm does not have a southern counterpart, however. 

    Since NGC 5597 has a disk companion, NGC 5595, to its north-west direction,
a first possibility for the origin of the northern spiral arm might be a tidal gravitational interaction \citep{com02}; NGC 5597 might be in an early phase of a merger with NGC 5595 as studied by computer simulations in major mergers using a hierarchical tree method to calculate gravitational forces and smoothed particle hydrodynamics to follow the evolution of gas \citep{mih96}. A second possibility might be an m=1 gravitational instability as has been studied by self-gravitating computer simulations \citep{jun96}. An m=1 instability might be confined between the center of the galaxy and its Outer Lindblad Resonance (OLR) radius \citep{jun96}, however, an m=1 instability has been observed mainly on the gas distribution (and not on the star distribution; see Figures 5, 13, 17, and 19 in \citealt{jun96}). A third possibility might be the result of disk self evolution with or without gas accretion \citep{bou02}, see for example their Figure 1 at 2.5 Gyrs, which roughly resembles the disk in NGC 5597. A fourth possibility would be the location near an OLR, where $\Omega_{\rm bar} = \Omega_{\rm gas} + \kappa/2$ \citep{sel93,but96,com02}. A final fifth possibility would be the location of an outer 4/1 (m=4) resonance (known as R1{$^\prime$} type 1 OLR subclass outer pseudoring) where $\Omega_{\rm bar} = \Omega_{\rm gas} + \kappa/4$ between corotation and OLR \citep{sel93,but96,com02}.

         Our VLA \hi\ emission results show a weak unresolved peak of cold gas at $\alpha \sim
14^h\,24^m\,28^s$, $\delta \sim -16\degr\,46'\,15\arcsec$, as if it were the center of a circular area ($\sim 15''$ radius) devoid of \hi\ gas. There is another area that lacks \hi\ emission located to the SW of the optical stellar bar at $\alpha \sim 14^h\,24^m\,26\rlap{.}{^s}2$, $\delta \sim -16\degr\,46'\,10\arcsec$. Furthermore, there is no atomic hydrogen cold gas emission from the position of the optical nucleus at $\alpha({\rm J2000}) = 14^h\,24^m\,27\rlap{.}{^s}49$, $\delta({\rm J2000}) = -16\degr\,45'\,45\farcs9$, nor from the circumnuclear region where there is hot gas ($T_{\rm e} \sim 10^4$\,K; \citealt{spi78}) as indicated by the H$\alpha$ emission \citep{gar96} and the 20\,cm radio continuum, most likely synchrotron, emission from that region (\citealt{con90}; Garcia-Barreto \& Momjian 2022 in preparation). There is also no \hi\ 21 cm emission from the south-west optical disk in what seems to be a large elongated hole with an approximate size of $\sim 17'' \times 7\farcs5$ at P.A.$\sim 110\degr$ EofN.

    As a comparison, many other disk galaxies generally show \hi\ 21 cm gas emission
well beyond the optical disk [i.e., among them M83, SBc(s) II \citep{rog74}, NGC 1300, SB(rs)bc \citep{eng89}, NGC 3783, SBa, \citep{gar99}, NGC 3147 S(rs)bc \citep{haa08}, Maffei 2 SAB(rs)cd \citep{hur96}]. In the case of gravitationally bound pairs, M51 is the best example with \hi\ 21 cm cold gas emission from inner regions that follows the spiral structures, but it shows an extended broad tail to larger radii to the east of M51 with projected length of about 90 kpc \citep{rot90}, while the M81 - M82 disk galaxy system, along with the small Irr II galaxy NGC 3077, shows \hi\ 21 cm cold gas emission from an extensive array of filamentary tidal structures threading all three galaxies and indicating their interaction \citep{cot76,van79,yun94}.

\subsubsection{VLA B-configuration \hi\ 21 cm Velocity Field and Kinematics}

    Our kinematical analysis was performed using the AIPS task {\it GAL}. After four
iterations, it fitted the \hi\ 21 cm parameters with the center at $\alpha({\rm J2000}) = 14^h\,24^m\,27\rlap{.}{^s}16$, $\delta({\rm J2000}) = -16\degr\,45'\,46\farcs64$, which agrees well with the photometric coordinates of the nucleus \citep{dia09}, the position angle of the semi-major axis with redshifted velocities P.A.$_{\rm red\,HI} \sim 100\degr$ EofN, the inclination of the disk $i \sim 36\degr$, and the systemic velocity $V_{\rm sys} \sim 2698$ km s$^{-1}$.  

    As noted above, the P.A. of the semi-major axis with redhsifted velocities is
$\sim 100\degr$ EofN, with the corresponding P.A. of zero velocities compared to the systemic velocity (minor axis) at $\sim 10\degr$ EofN\footnote{In a normal disk galaxy, in the approximation with only circular orbits ($v_R = 0, v_z=0$) the velocity field will be symmetric about the minor axis with one side showing redshifted velocities compared to the galaxy's systemic velocity, and the other side showing blueshifted velocities (see Fig. 8-17 in \citealt{mih81}).}. Thus, the north-east, east, south hemisphere shows the redshifted velocities while the north, north-west, west, south-west hemisphere shows the blueshifted velocities (figure of velocity field not shown, Garcia-Barreto \& Momjian 2022 in preparation). Taking the P.A.$\sim 100\degr$ EofN of the semi-major axis with redshifted velocities, and assuming the optical spiral arms are trailing, the hemisphere from north-west clockwise to south-east is closer to the observer. The orientation of the rotation axis vector points to the south-west at a projected P.A.$\sim 190\degr$ EofN. This orientation will be important when analyzing the \hi\ 21 cm gas kinematics and dynamics of both disk galaxies, NGC 5597 and NGC 5595, because the rotation axes are not parallel (Garcia-Barreto and Momjian 2022, in preparation).

    Figure 6 shows our \hi\ 21 cm VLA B-configuration spectrum of NGC 5597. The
overall shape is very similar to the spectrum from the single dish radio telescope in Nan\c{c}ay which is the only one that has an appropriate east-west angular resolution ($\sim 3\rlap{.}{'}6$) to isolate the \hi\ emission from NGC 5597 from that of NGC 5595 to the north-west \citep{pat03}. The \hi\ 21\,cm spectra from other single dish radio telescopes, such as Parkes (64\,m diameter) and Green Bank (91\,m diameter), included the emission from both disk galaxies because their angular resolution was poor ($\sim 11\rlap{.}{'}5$  and $\sim 8\rlap{.}{'}1$, respectively) \citep{mat92,spr05}. Thus, neither the Parkes, nor the Green Bank observations, could show the individual spectra of the two galaxies, and they could not accurately estimate the \added{systemic} \hi\ 21 cm velocity of either galaxy. The full widths of the \hi\ line in NGC 5597 at 50\% and 20\% of the peak, obtained with the VLA in B-configuration, are $\Delta V_{50\%} \sim 211.40$ km s$^{-1}$ and $\Delta V_{20\%} \sim 239.25$ km s$^{-1}$, respectively (see Figure 6). 
         
         An explanation for the low flux density seen in the velocity interval
from 2550 km s$^{-1}$ to 2590 km s$^{-1}$ is that it may be a plausible \hi\ 21 cm emission from a weak but extended low surface density gas at the south-west of NGC 5597, perhaps as a tidal filamentary structure at $\alpha \approx 14^h~24^m~26''$, $\delta \approx -16\degr 47'$ (Garcia-Barreto \& Momjian 2022 in preparation).

    In Figure 7, the left plot shows the fitted rotation curve with values slowly
rising from $V \sim 24$ km s$^{-1}$ at $R \sim 9\farcs23$ up to $V \sim 101$ km s$^{-1}$ at $R \sim 61\farcs8$. The curve was obtained in confocal circular anuli each of them $8\farcs0$ wide, and integrated from $R = 0\farcs0 \rightarrow 70\farcs0$ with both hemispheres \citep{rog74}. It shows a slowly rising curve, with a low velocity value near the central region. The right plot in Figure 7 shows the angular velocities of the \hi\ gas assuming circular orbits: $\Omega_{\rm gas}$ is shown by filled pentagons, $\Omega_{\rm gas} - \kappa(R) /2$ is shown by open circles \replaced{\footnote{See Appendix C for the origin of the epicycle frequency, $\kappa(R)$, and its mathematical expression.}}{\footnote{$\kappa(R)$ is the small radial motion of gas due to the non-axisymmetrical gravitational potential of a stellar bar. It may be estimated from the expression $\kappa(R) = \sqrt{4\Omega^2_{{\rm gas}} + 2R\Omega_{{\rm gas}} (d\Omega_{{\rm gas}})/dR}$}},  $\Omega_{\rm gas} + \kappa(R)/2$ is shown by open triangles, and $\Omega_{\rm gas} + \kappa(R)/4$ is shown by open squares. $\Omega_{\rm gas}$ shows a low value at R$ \sim 1.73$ kpc, then a high value at R$ \sim 2.4$ kpc and then a (normal) decreasing curve which is less pronounced than a simple $\Omega_{\rm gas} \propto 1/(R^{3/2})$, perhaps suggesting an extended central mass distribution\footnote{Modeling a detailed mass distribution in NGC 5597 is beyond the scope of the present study.}. The constant angular velocity corresponds to our estimated $\Omega_{\rm bar} \sim 15.3$ km s$^{-1}$ kpc$^{-1}$ assuming $\Omega_{\rm bar} = \Omega_{\rm gas}$ at corotation and $\mathcal{R} \equiv R_{\rm CR} / a_{\rm bar} =1$. The constant value of $\Omega_{\rm bar}$ crosses $\Omega_{\rm gas} + \kappa / 4$ at about a radius R$ \sim 6.73$ kpc (see section 3.1.5 for the details on the interpretation).

\subsubsection{VLA B-configuration: Total \hi\ 21 cm Neutral Atomic Gas Mass in NGC 5597}

    The total integrated \hi\ flux of NGC 5597 is $\int S_{\rm HI} dv \sim 2.9$ Jy km
s$^{-1}$, measured using the task {\it IRING} in AIPS with concentric rings, each $8\farcs0$ wide, from $R=0\farcs0 \rightarrow R=80\farcs0$. The total estimated atomic hydrogen mass of NGC 5597 is $M(\hi) \sim 1.02 \times 10^9$ M$_{\odot}$, which is very similar to \added{e.g.,} the \hi\ mass in Maffei 2 \citep{hur96}, a factor of 1.36 and 1.56 higher than the \hi\ masses of M82 and NGC 3077,  respectively, \added{and which are both disk galaxies in a minor merger or in gravitational interaction,} and is a factor of two lower than the \hi\ mass in M81\citep{yun94}.

\begin{figure*}
\includegraphics[width=14cm,height=12cm]{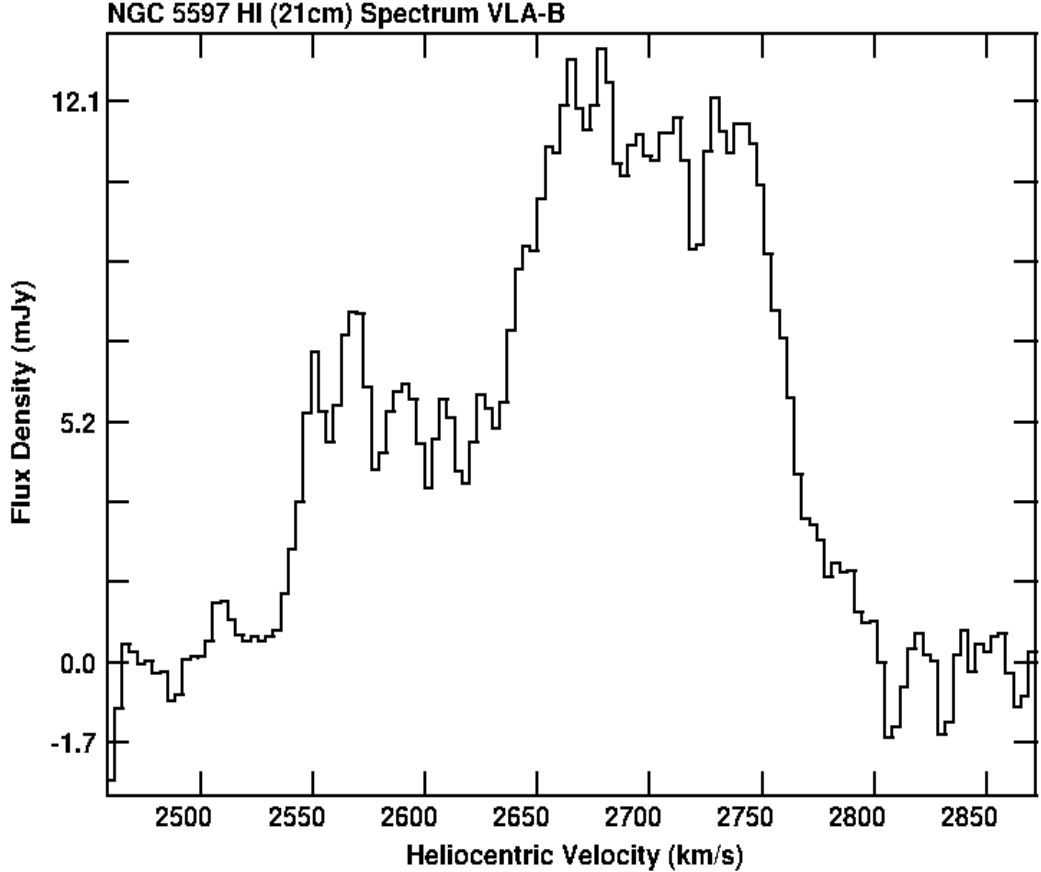}
\figcaption{VLA B-configuration \hi\ 21 cm spectrum of NGC 5597. The heliocentric systemic velocity, fitted by the task {\it GAL} in AIPS, is $V_{\rm sys\text{\textendash} helio} = 2698$ km s$^{-1}$, and the widths at 50\% and 20\% of the peak are $\Delta V_{50\%} \sim 211.40$ km s$^{-1}$ and $\Delta V_{20\%} \sim 239.25$ km s$^{-1}$, respectively. Comparing the flux density scale to previous Parkes (64 m) and Green Bank (91m) single dish radio telescope spectra is very difficult because their beams included both disk galaxies NGC 5595 and NGC 5597 \citep{mat92,spr05}. The shape of the VLA spectrum looks very similar to that of Nan{\c{c}}ay's, which has an elongated beam of $\sim 3\rlap{.}{'}6$ east-west $\times~22'$ north-south at FWHM \citep{pat03}.
\label{fig. 6}} 
\end{figure*}

\begin{figure*}
\includegraphics[width=9cm,height=9cm]{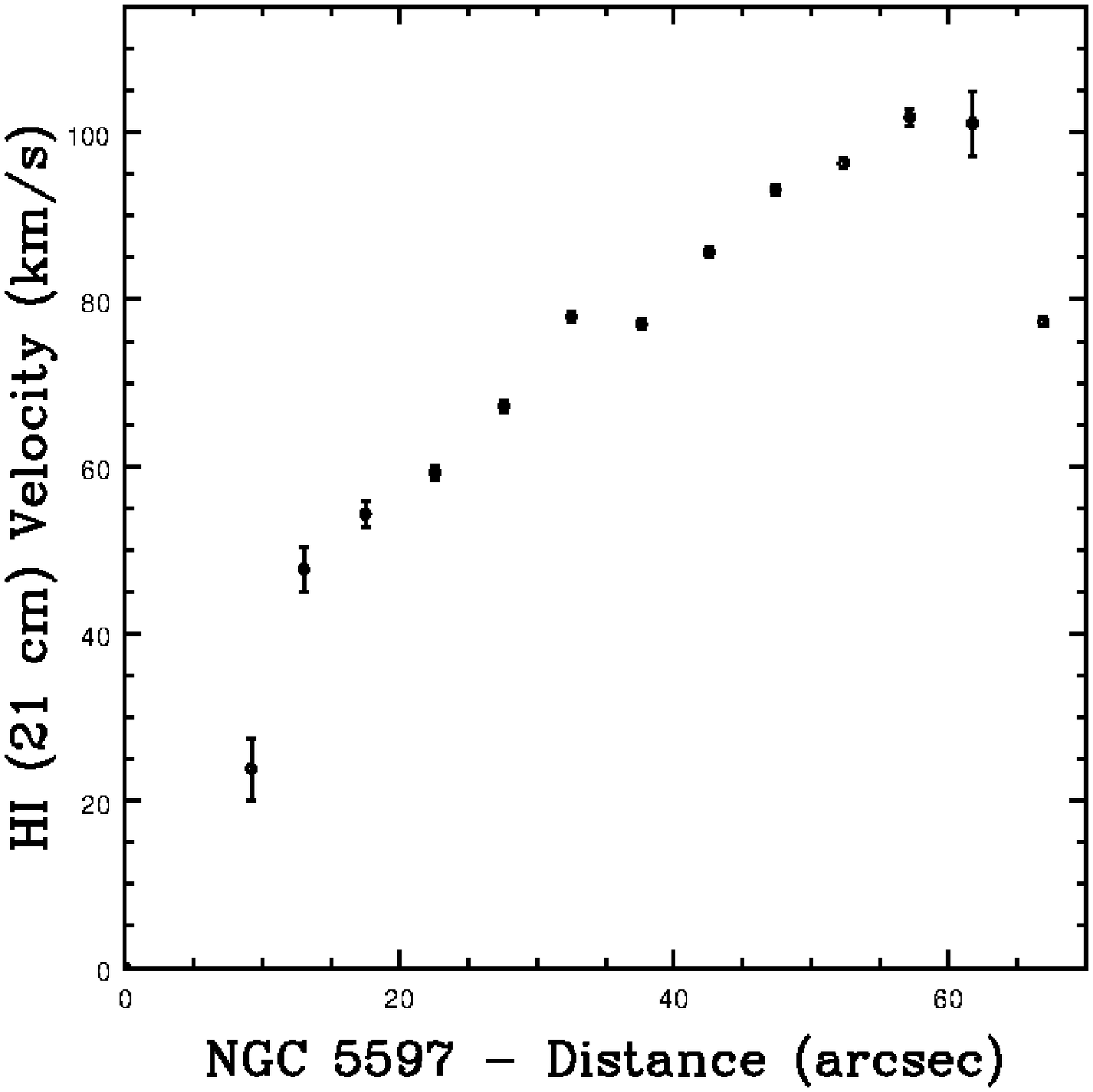}
\includegraphics[width=9cm,height=9cm]{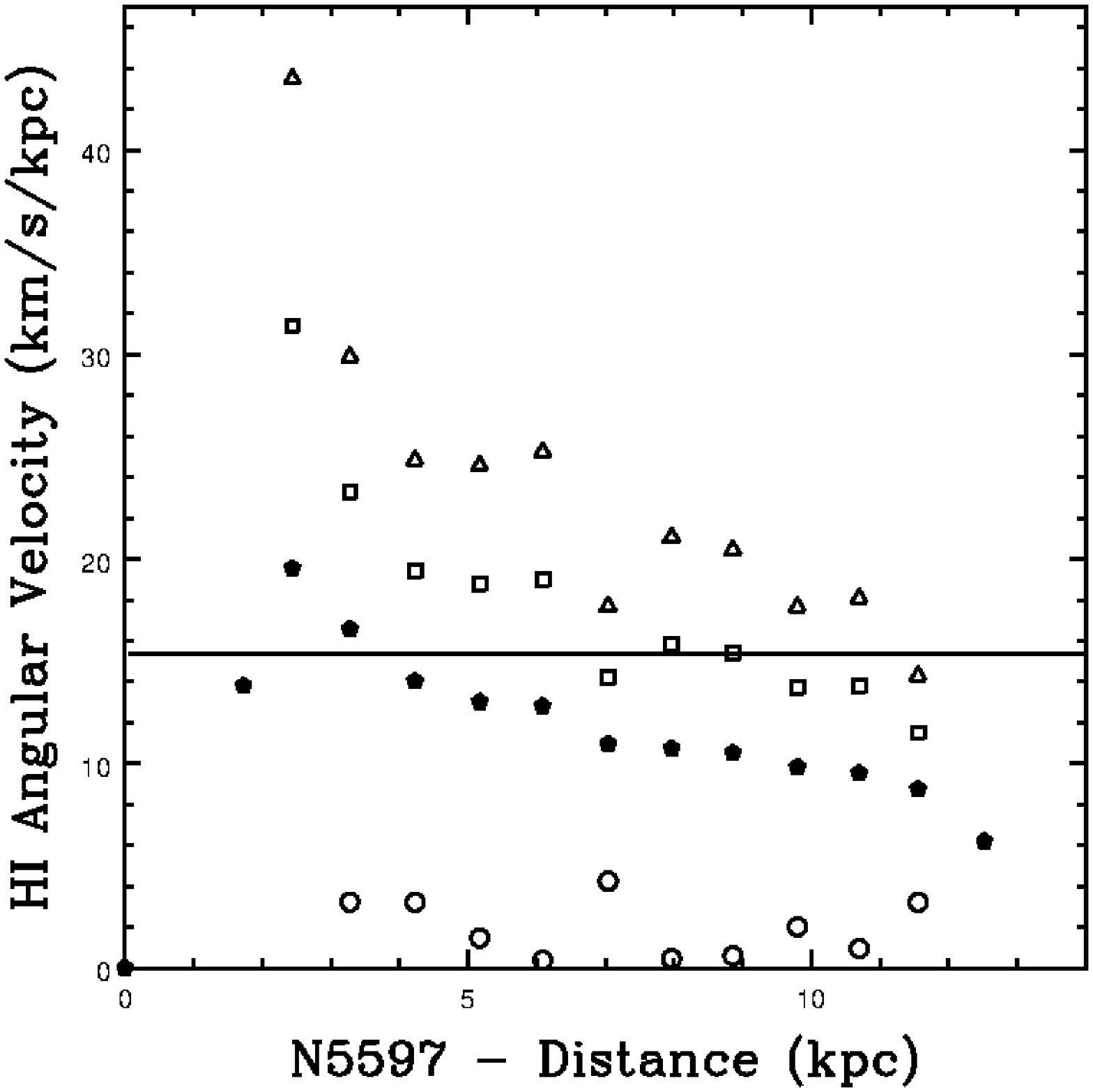}
\figcaption{{\it Left}: VLA B-configuration \hi\ 21 cm rotation curve of NGC 5597 after several iterations utilizing the task {\it GAL} in AIPS. The final fitted systemic heliocentric velocity is $V_{\rm sys\textendash helio} = 2698$ km s$^{-1}$. {\it Right}: The angular velocity of the  \hi\ 21 cm gas, $\Omega_{\rm gas}$ (filled pentagons), $\Omega_{\rm gas} - \kappa(R)/2$ (open circles),  $\Omega_{\rm gas} + \kappa(R)/2$ (open triangles), $\Omega_{\rm gas} + \kappa(R)/4$ (open squares), where $\kappa(R)$ is the epicycle frequency. The angular velocity pattern of a stellar bar, $\Omega_{\rm bar}$, is equal to the angular velocity of the gas, $\Omega_{\rm gas}$, in the epicycle approximation at corotation. Our estimation of $a^{r-ma}_{\rm bar} \sim 20''$ ($\sim 3.74$ kpc) in NGC 5597 (see Section 3.1.5) as the corotation radius would result in $\Omega_{\rm gas} \sim 15.3$ km s$^{-1}$ kpc$^{-1}$, therefore $\Omega_{\rm bar} \sim 15.3$ km s$^{-1}$ kpc$^{-1}$ (the constant horizontal line). This $\Omega_{\rm bar}$ value crosses $\Omega_{\rm gas} + \kappa(R)/4$ at a radius $R_{\rm Om=4}\sim 6.73$ kpc, and $\Omega_{\rm gas} + \kappa(R)/2$ at a radius $R_{\rm OLR} \sim 11.25$ kpc}. \label{fig. 7}
\end{figure*}

\subsubsection{Corotation radius, $R_{\rm CR}$, Angular Velocity Pattern $\Omega_{\rm bar}$, Outer Lindblad Resonance $R_{\rm OLR}$}

    As mentioned previously in subsection 3.1.1, we have six different estimates for $a_{\rm bar}$
in NGC 5597 {\it on the plane of the sky} and {\it projected on the kinematical major axis of the disk (ma)}.

     In this study, we think it would be more appropriate to analyze the six different estimates
for the length of the stellar bar in NGC 5597 ($a_{\rm bar}$) {\it projected on the ma of the disk} which are $a^{d-ma}_{\rm bar} \sim 1.01$ kpc, $a^{s-ma}_{\rm bar} \sim 1.74$ kpc, $a^{blue-ma}_{\rm bar} \sim 1.87$ kpc,  $a^{red-ma}_{\rm bar} \sim 3.74$ kpc, $a^{0.5b-ma}_{\rm bar} \sim 1.37$ kpc, and $a^{0.5r-ma}_{\rm bar} \sim 0.67$ kpc.

  Since we utilize the resonance method to estimate the angular velocity of the
stellar bar, $\Omega_{{\rm bar}}$, in NGC 5597 assuming that the corotation radius lies at the end of the bar's length, we proceed now to consider the different estimates of $a_{\rm bar}$ in combination with the kinematical \hi\ data, i.e., rotation curve and angular velocities (Figure 7), to estimate the radii of the outer $m=2$ and $m=4$ resonances.

    This method has been proven to be valid when applied to nearby bright galaxies
with ring structures near the ILR and the OLR \citep{bin87,com88,sel93,but96,elm96a,elm96b}. Assuming that the end of the boxy stellar bar is the radius of corotation, $R_{\rm CR}$  \citep{con80a,con80b,con81}, that is, $\mathcal{R} \equiv R_{\rm CR}/a_{\rm bar} = 1$, the resonance method, with an epicycle frequency $\kappa(R)$, indicates that at this radius $\Omega_{\rm gas} = \Omega_{\rm bar}$. 

    From the angular rotation curve (Figure 7--Right), we are obliged only to
consider values for $a_{\rm bar}$ (projected on the kinematical major axis of the disk) larger than 1.73 kpc (9$\farcs23$) because our observations did not detect any \hi\ 21 cm gas at smaller radii. Thus, we are left with only three possible values, namely, $a^{s-ma}_{\rm bar} \sim 1.74$ kpc, $a^{blue-ma}_{\rm bar} \sim 1.87$ kpc,  and $a^{red-ma}_{\rm bar} \sim 3.74$ kpc.

    First, if $a^{s-ma}_{\rm bar} \sim 1.74$ kpc ($\sim 9\farcs3$) were the corotation radius, then it would
result in $\Omega_{\rm gas}$ at an anomalous low value of $\sim 13.8$ km s$^{-1}$ kpc$^{-1}$ and it would set $\Omega_{\rm bar} = \Omega_{\rm gas}$ at that constant value that is well above the values of $\Omega_{\rm gas} - \kappa(R)/2$.
This constant value of $\Omega_{\rm bar}$ crosses  $\Omega_{\rm bar} \sim \Omega_{\rm gas} + \kappa(R)/4$ at a radius R$_{\rm Om=4} \sim 9.8$ kpc, and  $\Omega_{\rm bar} \sim \Omega_{\rm gas} + \kappa(R)/2$ at a radius $R_{\rm OLR} \sim 11.50$ kpc. 

     Second, if $a^{b-ma}_{\rm bar} \sim 1.87$ kpc ($\sim 10''$) were the corotation radius, then
it would indicate $\Omega_{\rm gas}$ at an anomalous low value $\sim 14.8$ km s$^{-1}$ kpc$^{-1}$ and it would set $\Omega_{\rm bar} = \Omega_{\rm gas}$ at that constant value that is also well above the values of $\Omega_{\rm gas} - \kappa(R)/2$. This constant value of $\Omega_{\rm bar}$ will cross $\Omega_{\rm gas} + \kappa(R)/4$ at a radius R$_{\rm Om=4} \sim 6.96$ kpc, and $\Omega_{\rm bar} \sim \Omega_{\rm gas} + \kappa(R)/2$ at a radius $R_{\rm OLR} \sim 11.37$\,kpc.

    Finally, taking $a^{r-ma}_{\rm bar} \sim3.74$ kpc ($\sim 20''$) as the
corotation radius, it would result in $\Omega_{\rm gas} \sim 15.3$ km s$^{-1}$ kpc$^{-1}$. It would then set $\Omega_{\rm bar} = \Omega_{\rm gas}$ at that constant value that is also well above the values of $\Omega_{\rm gas} - \kappa(R)/2$. This constant value of $\Omega_{\rm bar}$ will cross $\Omega_{\rm gas} + \kappa(R)/4$ at a radius R$_{\rm Om=4} \sim 6.73$ kpc, and $\Omega_{\rm bar} \sim \Omega_{\rm gas} + \kappa(R)/2$ at a radius $R_{\rm OLR} \sim 11.25$ kpc.

    Considering a) the extent, or the distance projected on the kinematical major axis of the
galaxy, of the red optical boxy isophotal contours (Figures 2 and 4) before the next isophotal contour that starts to extend azimuthally ($a^{r-ma}_{\rm bar} \sim 20''$, or $a^{r-ma}_{\rm bar} \sim 3.74$ kpc), b) the distance from the center to the mean optical red northern arm (R$^{nnw}_{\rm red} \sim 6.7$ kpc), c) the rotation curve and angular velocities from our \hi\ 21 cm results, and d) $\mathcal{R} \equiv {R_{\rm CR} /a_{\rm bar}} = 1$,  then our best estimate for $\Omega_{\rm bar}$ of the stellar bar in NGC 5597 with the resonance method is $\Omega_{\rm bar} \sim 15.3$ km s$^{-1}$ kpc$^{-1}$.

    How does this value of $\Omega^{\rm NGC 5597}_{\rm bar}$ compare with those obtained for
other galaxies? Excellent observational studies have estimated $\Omega_{\rm bar}$ for two barred galaxies: NGC 936 and NGC 4596. For NGC 936, $\Omega^{\rm NGC 936}_{\rm bar} \sim 60 \pm 14$ km s$^{-1}$ kpc$^{-1}$ \citep{mer95} derived from the velocity field of stars using the Tremaine-Weinberg (TW) method \citep{tre84}, $38 \leq \Omega^{\rm NGC 936}_{\rm bar} \leq 64$ km s$^{-1}$ kpc$^{-1}$ \citep{ken89}, and $\Omega^{\rm NGC 936}_{\rm bar} \sim 71$ km s$^{-1}$ kpc$^{-1}$ \citep{kor83}. 

    If we re-do the previously published analysis taking a) $a^{\rm NGC 936}_{\rm bar} \sim
51''$ {\it projected on the kinematical majoraxis of the galaxy}, which translates to $a^{\rm NGC 936}_{\rm bar} \sim 3.96$ kpc \citep{kor83,ken89,ken90}, b) $R^{\rm NGC 936}_{\rm CR} \sim 4.1$ kpc \citep{kor83},  c) a lower limit value of $\Omega^{\rm NGC 936}_{\rm bar} \sim 46$ km s$^{-1}$ kpc$^{-1}$ \citep{mer95}, then it would set $\mathcal{R} \equiv {R_{\rm CR} /a_{\rm bar}} \sim 0.96$ (a value very close to 1). 

    For NGC 4596, the values are $\Omega^{\rm NGC 4596}_{\rm bar} \sim 52 \pm 13$\,km\,s$^{-1}$
kpc$^{-1}$ \citep{ger99} using the TW method, and $\Omega^{\rm NGC 4596}_{\rm bar} \sim 43$\,km\,s$^{-1}$\,kpc$^{-1}$ \citep{ken90}. Also, there are two values reported for $a_{\rm bar}$, the first $a^{\rm NGC 4596}_{\rm bar} \sim 75''$ {\it projected on the kinematical major axis of the galaxy}, or $a^{\rm NGC 4596}_{\rm bar} \sim 5.7$ kpc \citep{kor79}, and the second $a^{\rm NGC 4596}_{\rm bar} \sim 66''$ {\it the radius where the intensity has fallen to half of its peak value along the bar's major axis}, or $a^{\rm NGC 4596}_{\rm bar} \sim 5$ kpc, \citep{ken90}. Additionally $R^{\rm NGC 4596}_{\rm CR} \sim 6.3$ kpc \citep{ken90}. Similarly, if we re-do the analysis for NGC 4596, taking a) the lower limit value of $\Omega^{\rm NGC 4596}_{\rm bar} \sim 39$ km s$^{-1}$ kpc$^{-1}$ \citep{ger99,ken90}, b) $a^{\rm NGC 4596}_{\rm bar}$ from Kormendy (1979), then it would suggest $\mathcal{R} \equiv {R_{\rm CR} /a_{\rm bar}} \sim 1$. If one were to take the upper limit value of $\Omega^{\rm NGC 4596}_{\rm bar} \sim 65$ km s$^{-1}$ kpc$^{-1}$ \citep{ger99,ken90} then it would suggest $\mathcal{R} \equiv {R_{\rm CR} /a_{\rm bar}} \sim 1.1$\footnote{If, however, one were to take $a^{\rm NGC 4596}_{\rm bar}$ from Kent (1990) together with the lower limit value of $\Omega_{\rm bar} \sim 39$ km s$^{-1}$ kpc$^{-1}$ \citep{ger99}, then it would suggest $\mathcal{R} \equiv {R_{\rm CR} /a_{\rm bar}} \sim 1.15$, and if one were to take the upper limit value $\Omega_{\rm bar} \sim 65$ km s$^{-1}$ kpc$^{-1}$ \citep{ger99} then it would suggest $\mathcal{R} \equiv {R_{\rm CR} /a_{\rm bar}} \sim 1.26$}. 
    
    Therefore, we believe that our assumption of $\mathcal{R} \equiv {R_{\rm CR}
/a_{\rm bar}} \sim 1$ is reasonable, and our value of $\Omega_{{\rm bar}}$ in NGC 5597 SBc(s), estimated using the resonance method, is similar or a bit smaller than the values for other bright and nearby barred galaxies, i.e., $\Omega_{{\rm bar}} \sim 60$\,km\,s$^{-1}$\,kpc$^{-1}$ in NGC 1326 (RSBa with a circumnuclear ring) \citep{gar91b}, $\Omega_{{\rm bar}} \sim 134$\,km\,s$^{-1}$\,kpc$^{-1}$ in NGC 1415 (SBa with a geyser or low velocity bipolar nuclear H$\alpha$ outflow) \citep{gar19}, $\Omega_{{\rm bar}} \sim 39 - 43$ km s$^{-1}$ kpc$^{-1}$ in NGC 3367 (SBc compact nuclear ring and radio bipolar outflow) \citep{gar01,gar14}, $\Omega_{{\rm bar}} \sim 38$ km s$^{-1}$ kpc$^{-1}$ in NGC 3783 (SBa Seyfert 1 and strong X-ray emission and X-ray variability) \citep{gar99}, and $\Omega_{{\rm bar}} \sim 36$\,km\,s$^{-1}$\,kpc$^{-1}$ in NGC 4314 (SBa(rs)p with a circumnuclear ring) \citep{gar91a}.

    Lastly, from the right plot in Figure 7, $\Omega_{{\rm bar}}$ crosses $\Omega_{{\rm gas}} +
\kappa(R)/4$ at a radius, $R \sim 6.73$ kpc; this radius corresponds to the spatial location of the north optical spiral arm structure seen in Figures 1 and 2. From our \hi\ 21 cm VLA B-configuration kinematical observations we cannot estimate any radius for a plausible ILR. Furthermore, our observations combined with the resonance method and $\mathcal R =1$ suggests that a plausible interpretation for the north spiral optical arm might be an outer m=4 resonance in NGC 5597 \citep{con89,ath92}.

\section{Summary and Conclusions}

In this study, we have obtained VLA B-configuration \hi\ 21 cm kinematical data from the disk galaxy NGC 5597 SBc(s). We were able to obtain the two dimensional velocity field and to estimate the {\it best fit} rotation curve assuming the neutral gas is in circular orbits around the nucleus. From the rotation curve of NGC 5597, we have calculated $\Omega_{{\rm gas}}$, $\kappa({\rm R})$, $\Omega_{{\rm gas}} - \kappa({\rm R})/2$, $\Omega_{{\rm gas}} + \kappa({\rm R})/2$ and $\Omega_{{\rm gas}} + \kappa({\rm R})/4$. 

    From our previously published studies, we have our best estimate for $a_{\rm bar}$ as
$a^{r-ma}_{\rm bar} \sim 20''$ {\it projected on the kinematical major axis of the \hi\ cm in the disk}, or $a_{\rm bar} \sim 3.74$ kpc.

    We have assumed that the radius at which corotation occurs, $R_{{\rm CR}}$, is just at the
end of the semimajor axis of the stellar bar in NGC 5597, $\mathcal{R} \equiv R_{{\rm CR}}/a_{{\rm bar}} = 1$ \citep{con80a,con80b,kor83}. With this assumption and the angular velocities, we estimated the angular velocity for the stellar bar in NGC 5597 to be $\Omega_{\rm bar} \sim 15.3$\,km\,s$^{-1}$\,kpc$^{-1}$. Additionally, this constant value of $\Omega_{{\rm bar}}$ crosses the curve $\Omega_{{\rm gas}} + \kappa(R)/4$ at 6.73 kpc, which is very similar to the radius of the spatial location of the optical north spiral arm in NGC 5597. Therefore, we have interpreted this as an outer m=4 resonance arm.

\begin{acknowledgements}
We thank the anonymous referee for suggestions and comments that
have significantly enhanced the content of this manuscript.
\end{acknowledgements}

\vspace{5mm}
\facilities{EVLA}

\software{
AIPS: \citep{G2003},
CASA \citep{mcm07}
}

\end{document}